# How Microplastics cross the Buoyancy Barrier: A multi-scale Study

**Thomas Witzmann**[1], Anja F. R. M. Ramsperger[2, 3], Hao Liu[4], Yifan Lu[5], Holger Schmalz[6], Lucas Kurzweg[7, 8], Tom C. D. Börner[7], Kathrin Harre[7], Andreas Greiner[6], Christian Laforsch[2], Holger Kress[3], Christina Bogner[5], Stephan Gekle[6], Andreas Fery*[1, 7], and Günter K. Auernhammer*[1]

[1] *Leibniz-Institute of Polymer Research Dresden e.V., Physical Chemistry and Polymer Physics, Hohe Str. 6, 01069 Dresden, Germany*

[2] *Animal Ecology I and BayCEER, University of Bayreuth, 95447 Bayreuth, Germany*

[3] *Biological Physics, University of Bayreuth, 95447 Bayreuth, Germany*

[4] *Biofluid Simulation and Modeling, University of Bayreuth, 95447 Bayreuth, Germany*

[5] *Ecosystem Research, Institute of Geography, Faculty of Mathematics and Natural Sciences, University of Cologne, Zülpicher Straße 45, 50674 Cologne, Germany*

[6] *Macromolecular Chemistry and Bavarian Polymer Institute, University of Bayreuth, 95447 Bayreuth, Germany*

[7] *Physical Chemistry of Polymeric Materials, Technische Universität Dresden, Hohe Str. 6, 01069 Dresden, Germany*

[8] *Faculty of Agriculture, Environment and Chemistry, University of Applied Sciences Dresden, Friedrich-List-Platz 1, 01069 Dresden, Germany*

Corresponding authors: fery@ipfdd.de, auernhammer@ipfdd.de

## Abstract

Microplastics (MPs), though less dense than water, are frequently recovered from sediments in aqueous environments, indicating they can cross the buoyancy barrier. We quantify eco-corona mediated MP-sediment attraction and MP transport from the nanoscale to the macroscale, linking all scales to a coherent mechanism explaining how MP overcome buoyancy and settle in sediments through interaction with suspended sediment.

Colloidal probe atomic force microscopy (CP-AFM) detected attractive forces (0.15 – 17 mN/m) enabling heteroaggregation. Microscale tests confirmed aggregation and on larger scales sediment retention more than doubled with an eco-corona. Simulations showed that environmental shear force ($4 \cdot 10^{-4}$ mN/m) cannot disrupt aggregates. In sedimentation columns, biofilm-covered MPs settled twice as often as plain MPs in bentonite suspensions. MP retention increased by 32 %. These results demonstrate that eco-corona/biofilm-mediated heteroaggregation is a robust pathway for MP sinking, accumulation, and retention in sediment beds. By identifying physical interaction thresholds and aggregation stability, we provide mechanistic insight into MP fate, highlight probable accumulation hotspots, and offer an evidence base for improved risk assessment and environmental modelling.

# Introduction

Microplastics (MPs) have been recovered from the deepest ocean trenches, buried in riverbed sediments, and trapped in centuries-old lake cores—despite being less dense than water.[1] Such observations challenge the intuitive view that buoyant particles should remain afloat. MPs are buoyant due to their low density, hydrophobicity, and occasional attachment of air bubbles, yet they routinely cross the buoyancy barrier and settle into sediments. [2-4] This paradox points to processes that both disperse MPs in the water column and drive their sinking, enabling subsequent transport and retention in sediments.

Beyond their puzzling transport behaviour, MPs are a rising global concern for human and environmental health.[5-7] With sizes from 1 µm to 1 mm[8], they possess a high surface-area-to-volume ratio, enhancing sorption of persistent organic pollutants[9], heavy metals[10,11], antibiotics[12,13], and other contaminants[14,15]. Preventing their negative impacts and enabling effective removal requires understanding of the environmental conditions and mechanisms—accumulation[16], aggregation and sedimentation[17]—through which MPs interact with their surroundings. Yet, previous studies have examined these processes in isolation, leaving the coupled mechanisms that govern MP fate across scales poorly understood. [17-22]

Besides particles, there are microorganisms like bacteria and algae and different kinds of natural organic matter (NOM) abundant in the environment, all of which adsorb on the MP surface and alter the interaction of MP particles. For MP particles in the same size range as most bacteria (1 – 10 µm), bacteria are unable to adhere to the surface and thus cannot form a biofilm. However, due to continuous degradation and fragmentation the majority of MP particles is in or below that size range.[23,24] Therefore, when sub-10 µm MP particles enter the aquatic environment an eco-corona will form on the surface rather than a biofilm involving whole microorganisms.[25] The eco-corona is composed of adsorbed proteins, extracellular polymeric substances (EPS) and NOM like humic substances forming a soft layer on the surface. Eco-corona formation not only alters the MP surface properties[26,27] but also influences the interaction[28,29] of MP particles with other particles and even with cells[30,31]. The formation and properties of the eco-corona are thereby influenced by the type and concentration of adsorbing macromolecules like different NOMs,[32-35] salt type and concentration,[36] pH, temperature which vary across river,[37] lake, soil[38] and saltwater[39,40]. In addition, the physicochemical properties [32,35,41] of the MP particle, such as chemistry, charge, polarity, hydrophobicity play a crucial role in eco-corona formation.[33,42-44]

In natural waters, there are different kinds of suspended sediment particles present. Most often sediment is referred to inorganic materials such as quartz, clay minerals or calcium carbonate from shells present in the water column. Sediment originates e.g. from eroded rocks and soils or weathering upon thermal stress within rocks. During rain fall or flooding events the material is suspended and transported to the aqueous environment. There, the abundance of suspended sediment generally exceeds the abundance of MP particles. Therefore, MP particles are more likely to interact with suspended sediment rather than interacting with other MP particles. To date, several studies have investigated the heteroaggregation behaviour or interaction between suspended sediment and MP particles. They primarily focused on the influence of particle type, particle number ratio (or concentration ratios), salt concentration and composition, and pH on the aggregation process.[17,45-50]

Since the presence of macromolecules can alter the interaction of MP with natural particles, it will also affect MP transport in the environment. However, most MP transport studies focus mainly on the MP size, shape, quantity and density.[51] Rarely has MP transport been linked to the MP surface properties and their interaction with sediment particles[52-54] or the presence of natural macromolecules[20,55,56]. The combined effects of these factors on MP transport remain poorly understood, especially across multiple scales. This means, the mechanisms governing heteroaggregation of eco-corona- and biofilm-covered MPs in the presence of sediment particles are still underexplored. Better insights into these processes may help to understand how MPs overcome the buoyancy barrier and to predict MP behaviour in the environment based on a general mechanism.

To close these gaps, we present for the first time a quantitative, multi-scale mechanism showing that eco-corona and biofilm mediated interaction influence the buoyancy of MPs and their transport in the aqueous environment. At the nanoscale, we use colloidal probe atomic force microscopy (CP-AFM) to quantify the interaction forces between individual MP and model sediment ($SiO_2$) particles under different ionic strengths and eco-corona coverage levels (Figure 1A). Then, we test the heteroaggregates stability under environmental relevant shear flow conditions with Lattice-Boltzmann simulations.[57,58] We link the prevailing interaction mechanism to larger scales and provide reasoning for MP transport. On the microscale we demonstrate the influence of the eco-corona on single aggregates and their transport behaviour (Figure 1C). We evaluate whether the heteroaggregates could persist under typical environmental conditions. On the macroscale we also show through sedimentation experiments that MP particles although less dense than water sediment after heteroaggregation with suspended sediment (Figure 1D). Furthermore, we demonstrate that MP particle mobility in sediment is reduced due to eco-corona mediated inter-particle attraction and we quantify the drag force within sediment pores under water flow with simulations (Figure 1E). With this we provide a mechanistic understanding on why MPs are able to cross the buoyancy barrier and accumulate in and at sediment beds.

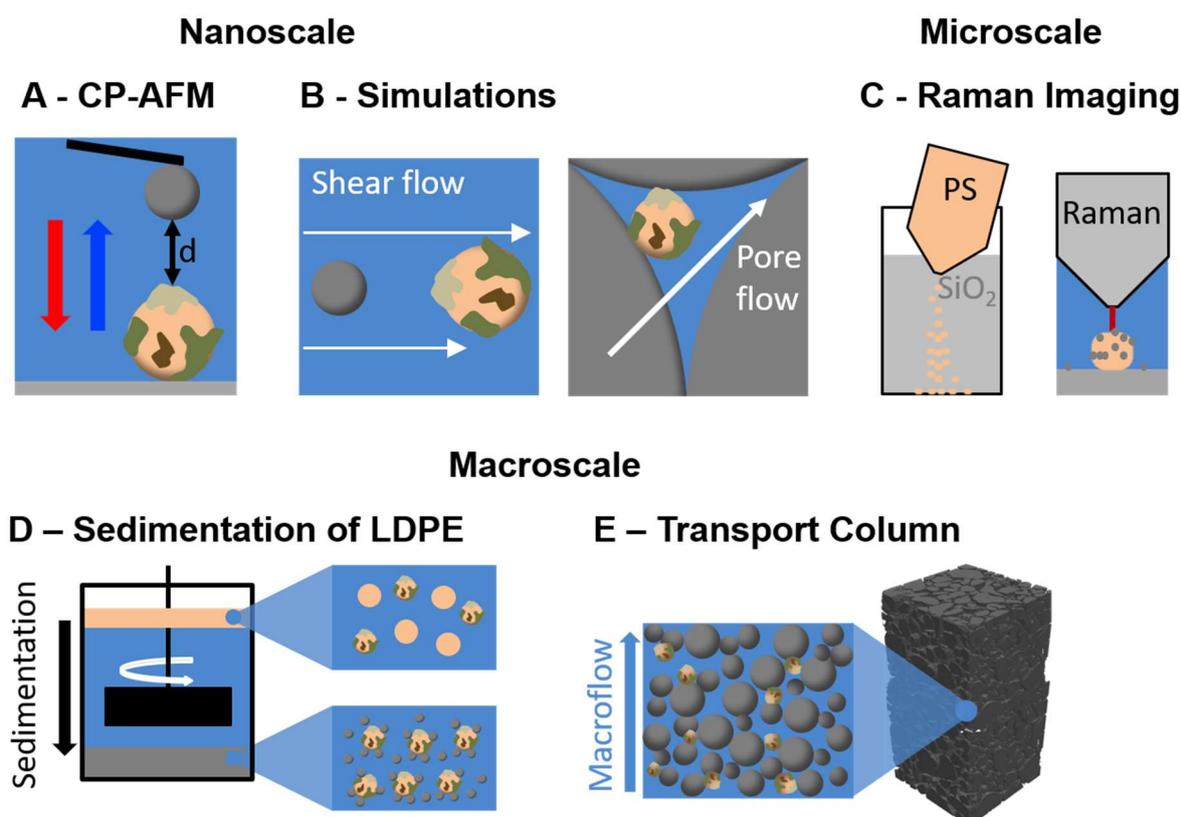

*Figure 1: Overview of experimental methods applied in this work. (A) CP-AFM measurements in KCl solutions of different concentration between a model sediment ($SiO_2$, dark grey) colloidal probe and PS particle (orange) to investigate the eco-corona mediated interaction forces. (B) Simulation of the separational force in a turbulent flow and in a laminar flow in porous media between model sediment and MP particle to assess the stability of heteroaggregates under flow. (C) Aggregation experiments of plain and eco-corona covered PS particles with $SiO_2$ followed by Raman imaging for aggregate characterisation. (D) Sedimentation of LDPE particles in the presence and absence of bentonite and biofilm. (E) Transport column experiments with model sediment. Suspensions of plain and eco-corona covered PS particles in 1 mM KCl were pumped through the columns.*

# Eco-corona mediated attraction on different length scales

## Nanoscale: Eco-corona mediates heteroaggregation between MP and sediment

To gain fundamental insights on why the eco-corona mediates the sinking of MP particles, we investigated the individual particle–particle interactions between eco-corona covered MPs and sediment particles. CP-AFM was used to study the interaction of polystyrene (PS) particles covered with an eco-corona and silica-sediment ($SiO_2$) particles. Details on CP-AFM measurements and evaluation can be found in Materials and Methods section and SI (Figure S11)

CP-AFM detected long ranged negative, i.e. attractive forces, for the interaction between sediment particle and MP covered with an eco-corona. In contrast, there are no attractive forces present for plain MP without an eco-corona which we published elsewhere.[28] The long ranged attractive forces are displayed in the force-separation curves of Figure 2A. Generally, the eco-corona mediated attraction is stronger for the retraction (blue) compared to the approach (red) of particles. We refer to this mechanism of attraction in the following as eco-corona bridging.[59] Eco-corona bridging can be induced by various mechanisms like charge attraction, van-der-Waals forces or hydrogen bonds. They occur between adsorbed macromolecules of the eco-corona and free adsorption sites at the sediment particle. The light-coloured area gives the range of bridging in terms of force and separation distance measured in our experiments. The range of the bridging force on approach ($x_3$ in Figure 2A) exceeds 100 nm. This is more than three times the maximal possible Debye length in our experiments (see Figure S1). The Debye length describes the electrostatic interaction length and is influenced by the ionic strength. Hence, the attraction is not based on electrostatics, but macromolecular interaction.

MP without an eco-corona showed no bridging, supporting the role of the eco-corona. Eco-corona bridging is the key mechanism for the spontaneous formation of heteroaggregates between MP and sediment particles.

## Nanoscale: Eco-corona bridging exceeds simulated hydrodynamic separational force

In the aqueous environment MP-sediment heteroaggregates are exposed to currents of the flowing water. We test whether the shear flow in water is able to disintegrate these heteroaggregates, by comparing the eco-corona bridging forces to simulated hydrodynamic forces. With these Lattice-Boltzmann simulations we find a shear flow induced separation force of approximately *2.5 · 10$^{-13}$ N*. With the effective radius (eq. (5)) $R_{eff} \approx 0.56~\mu m$ this yields *F/$R_{eff}$ ≈ 4 · 10$^{-4}$ mN/m* which is smaller than any bridging force measured with CP-AFM. More precisely, the eco-corona bridging force is by a factor of $10^3$ to $10^4$ stronger than shear flow induced separation forces. Thus, typical shear flows present in the aqueous environment are generally neither able to prevent eco-corona mediated heteroaggregation by bridging nor can they disintegrate these heteroaggregates.

The results show that eco-corona bridging forces are sufficient to maintain stable MP-sediment heteroaggregates under typical environmental shear flows. Thus, enabling long-term persistence in aquatic systems.

## Microscale: Eco-corona bridging mediated heteroaggregation and sedimentation

On the microscale, we test if the attachment of sediment particles to MP increases the overall density leading to sedimentation. We infused PS dispersions into a concentrated $SiO_2$ particle dispersion to foster sediment attachment.

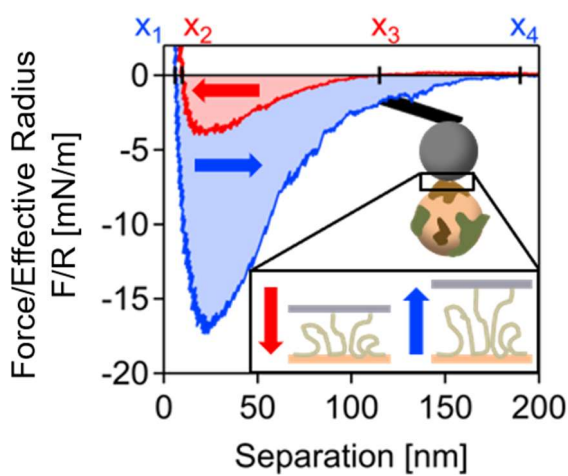
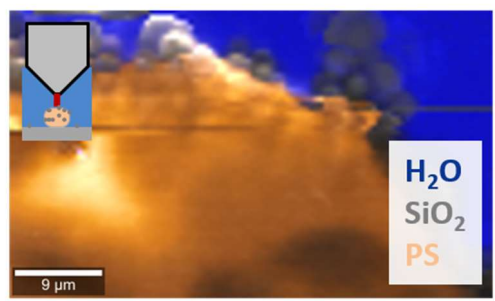
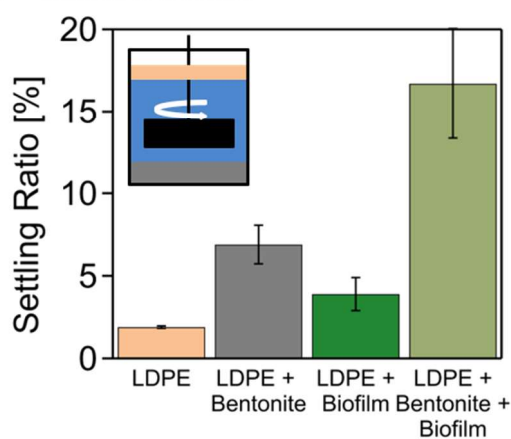
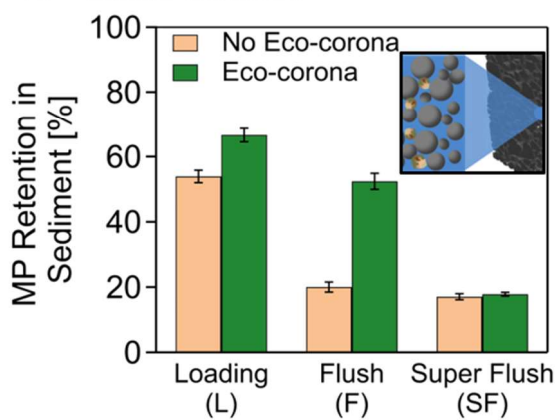

Figure 2: Eco-corona bridging forces across different length scales. Nanoscale: **(A)** Eco-corona covered PS particles (orange) bind to $SiO_2$ particles (grey) via eco-corona bridging forces. Bridging forces in force-separation curves during approach (red) and retraction (blue) of the particles. Once formed these aggregates are very stable (blue). Light coloured areas display the range of bridging forces found within CP-AFM experiments. $x_1$ and $x_2$ refer to the repulsion range on retraction and approach, respectively. $x_3$ and $x_4$ are the range of attractive forces on approach and retraction, respectively i.e. bridging ranges. Microscale: **(B)** Bridging forces result in PS-$SiO_2$ aggregation and sedimentation. The spatial component distribution extracted from confocal Raman imaging shows $SiO_2$ particles (grey) attached to PS particles with eco-corona (orange) in water ($H_2O$, blue). Macroscale: **(C)** Comparison of settling ratios of plain and biofilm covered LDPE particles after aggregation-sedimentation experiments (n = 6) in river water or bentonite suspension. Plain LDPE in river water has a settling ratio of $1.9 \pm 0.1\,\%$. In the presence of bentonite or biofilm the settling ratios increase to $6.9 \pm 1.2\,\%$ and $3.9 \pm 1.0\,\%$, respectively. The combined effects of bentonite and biofilm increase the settling ratio to $16.7 \pm 3.4\,\%$. Results for LDPE particles in the presence of biofilm and bentonite are in line with the observations for eco-corona bridging.; **(D)** In transport column experiments, eco-corona covered PS particles show a significantly higher retention after loading (L) $67 \pm 2\,\%$ and flush (F) $52 \pm 2\,\%$ compared to plain MP $20 \pm 2\,\%$ and $54 \pm 2\,\%$, respectivley. After super flush (SF) the difference in retention is not significant.

The eco-corona enables the suspended sediment to attach on all sides of the MP particle, inducing sedimentation (Figure S3). In contrast, plain MP particles without an eco-corona have no sediment particles attached and float on top of the water (Figure S5). Confocal Raman measurements at distinct spots show that the aggregates consist of different materials indicating the eco-corona bridging of big MP and small sediment particles (Figure S4). Figure 2B displays the spatial component distribution of such a heteroaggregate extracted from Raman imaging. The MP particle is displayed in orange and the sediment particles in grey, clearly confirming the formation of heteroaggregates in water. The heteroaggregates formed by eco-corona bridging were stable even after several hours of measuring time withstanding the possibility for sediment to settle. The sediment shows a higher co-localisation to the MP as shown in the optical microscopy images taken at different z-heights above the substrate in Figure S3. These results demonstrate that the eco-corona mediates heteroaggregation by bridging forces. The attachment of suspended sediment particles with much higher density compared to MP increases the overall density of the aggregate. This enables MP-sediment heteroaggregates to overcome the buoyancy barrier and subsequently settle in the environment. Thereby, removing MP from the water column. However, many toxicity tests on an organism level are done with e.g. daphnia in the water column.[60,61] Instead, more studies should be done on sediment-associated organisms, because MP particles sediment.[5,62]

Our results point toward a general mechanism, which also occurs for low-density MP, as we will discuss in the next section.[52,53,63]

## Macroscale: Biofilm mediates sedimentation of buoyant LDPE

In sediment experiments, we test if the eco-corona bridging mechanism can be also observed for biofilms. We therefore, use bigger MP particles with a size of 63 – 200 µm to allow microorganismal adhesion. Furthermore, we use LDPE and bentonite as a different MP and sediment material, respectively, to show the wide applicability of the proposed mechanism.

The sole presence of LDPE in river water leads to a settling ratio of 2 %. In the presence of bentonite or biofilm the settling ratio of LDPE increases to 7 % and 4 %, respectively. The combined effects of bentonite and biofilm significantly increase the settling ratio by 15 % from 2 % to 17 % (Figure 2D). Through biofilm formation, LDPE becomes more hydrophilic and the overall density increases (see Figure S6). Thereby, LDPE particles are dispersed in the water column, making them available for heteroaggregation with suspended sediment. Through the biofilm mediated attachment of the bentonite the aggregate's overall density overcomes the buoyancy barrier of LDPE, resulting in increased sedimentation.

Molazadeh et al. found that turbulent flows are able to transport PE particles vertically in the water column and close to the sediment.[19] Zones with high quantities of suspended sediment particles foster heteroaggregation with MP. The heteroaggregates will sediment especially in slow moving waters. Here, flow velocities and hence shear rates are small and will not disintegrate the heteroaggregates or resuspend them in the water column. This is supported by sediment analyses along the Elbe River, which show a higher abundance of MPs in areas of slow or stagnant flow compared to those of fast-flowing water.[64] The aforementioned characteristics of water, i.e. zones with high quantities of suspended sediment followed by zones of low flow velocity aid, in combination with MP entry pathways, MP hotspot analysis. Furthermore, these characteristics can be used as prerequisite for exposure risk-based toxicity tests

In this experiment, LDPE acts as a more critical example to the effects of the heteroaggregation induced sedimentation, since the density of LDPE (0.91 -0.94 g/cm³) is lower than water and compared to PS (1.05 g/cm³). In addition, it is very difficult to produce well-defined MP by grinding of LDPE due to its comparably soft nature (in comparison to PS), mostly particles are strongly deformed and size distribution is broad. This might alter significantly results from column transport, discussed in the next section, as the rough structure might promote fixation of the MP in the column (compared to a more spherical shape). Hence, LDPE MP was studied due to its lower density but all other studies were conducted with the more well-defined PS MP. Nevertheless, the results are in line with the eco-corona bridging induced heteroaggregation and sedimentation mechanism. The buoyancy barrier is much higher for MPs of lower density. We were able to prove that even such a high barrier can be overcome by biofilm mediated aggregation. The proposed heteroaggregation and sedimentation mechanism seems to be independent of MP polymer type, size and shape, and sediment particle chemistry. Eco-corona and biofilm mediate crossing the buoyancy barrier and sedimentation even though the plain MP would float on top of the water.

## Macroscale: Eco-corona enhances MP retention in sediments

In natural conditions, once MPs settle at the sediment, water continues to flow through the sediment–MP matrix, potentially transporting MPs along with it.[65] Column transport experiments were performed to test if the strong eco-corona bridging forces would also influence the retention of MP in the sediment. We therefore, infuse dispersions of plain and eco-corona covered PS MP particles into different columns containing quartz sand ($d_{50}$ = 390.8 µm).

After loading, 67 ± 2 % of eco-corona covered and 54 ± 2 % of plain MP particles were retained, respectively (p < 0.005, Figure 2D). Thereafter, the columns were flushed with background solution (0.08 mL/min) and the retention remained high at 52 ± 2 % for eco-corona covered MP, but dropped to 20 ± 2 % for plain MP (p < 0.001). An exhaustive statistical analysis is provided in Figure S8. The more than twofold increase in retention in the presence of the eco-corona indicates that the separational forces present at a flow rate of 0.08 mL/min within the column are not strong enough to disrupt the eco-corona-bridging forces between MP and sediment particles. This supports the idea of the eco-corona bridging mechanism, which promotes MP retention and accumulation in sediment beds.[66] He et al. also observed an increase in retention of carboxylated-PS MP in column transport experiments where quartz sediment was biofilm covered.[67] When the flush rate was increased from 0.08 to 0.25 mL/min to mimic stronger flow rates, MP retention was reduced to 17 – 18 % for plain and eco-corona covered MP (Figure 2D and Figure S9). The average pore flow velocity $u_{avg}$ was calculated for flush (0.08 mL/min) and super flush (0.25 mL/min) conditions to be $u_{0.08}$ = 3.56 · 10$^{-5}$ m/s and $u_{0.25}$ = 1.11 · 10$^{-4}$ m/s, respectively. These values are within the range observed in natural riverbed infiltration (10$^{-5}$ to 10$^{-3}$ m/s)[68,69], allowing the experiments to realistically reproduce MP transport and retention under environmentally relevant hydraulic regimes. This suggests that only high flow rates (with high separational forces) can overcome the eco-corona bridging forces. The remaining MP particles are mostly mechanically trapped within the sediment, which depends on pore and MP particle size. While this retention largely prevents the immediate return of MP particles to the water column, these trapped MPs may gradually fragment or undergo surface alterations due to various weathering processes, potentially enabling further transport into deeper sediment layers or downstream under high flow. Importantly, these smaller, surface-modified MPs are more susceptible to eco-corona formation, reinforcing particle–sediment interactions and promoting further aggregation and long-term retention. This implies that even low-density or originally buoyant MPs can ultimately overcome the buoyancy barrier and accumulate in sediments, potentially serving as a long-term environmental sink.

Having shown that eco-corona bridging more than doubles the retention of MP in sediment, we estimated the shear rate acting on the MP in the sediment pores and explored how geometry influences the separational force between MP and sediment particles. Assuming a monodisperse sediment particle size of 390.6 µm, the pore radius (a) is 30 µm. Inserting in Hagen-Poiseuille´s law:

$$u = u_{max}(1 - \frac{r^2}{a^2}) \qquad (1)$$

$$u_{max} = 2\, u_{avg} \qquad (2)$$

with $u$ the flow velocity, $u_{max}$ maximum flow velocity, $r$ particle radius and $u_{avg}$ we can calculate the shear rate acting on a MP particle within a pore after differentiating $u$ to $r$ and set $r = a$:

$$\dot{\gamma} = \frac{4 u_{avg}}{a} \qquad (3)$$

We obtain a shear rate of $\dot{\gamma}_{0.08}$ = 4.9 s$^{-1}$ for the flush and $\dot{\gamma}_{0.25}$ = 15.1 s$^{-1}$ for the super flush condition. The shear rate for flush (4.9 s$^{-1}$) is similar to the shear rates found in rivers (3.1 s$^{-1}$), which we have already discussed, is 3 to 4 orders of magnitude lower than necessary to overcome eco-corona bridging forces. Pore flow simulations showed that only at extremely high pore flow velocities ($u_{p,\,sim}$ = 1 m/s, 9.4 – 29.3 mN/m, Figure S7) eco-corona bridging forces (4 – 17 mN/m) are disrupted. The simulations in Figure S7 further show that the detailed pore geometry has a minor influence on the retention of MP in sediment as long as the pore flow velocity stays constant. However, it has to be considered that the separational forces, between sediment and MP in pores, act in a different direction to CP-AFM experiments. This explains why only a three-fold increase in shear rate is able to remove substantially more MP from the column than the comparison of force values would suggest.

The relative particle velocity, i.e. the difference in particle velocity between MP and sediment, also influences the eco-corona bridging force (Figure S10). The bridging force increases with lower relative particle velocity and vice versa, because at slow relative particle velocities macromolecules of the eco-corona have more time to relax and find optimal adsorption sites, thereby increasing the bridging force. Once heteroaggregates have formed the relative particle velocity has a negligible effect on bridging.

Our findings reveal that eco-corona formation generates macromolecular bridging forces that exceed typical environmental shear, enabling stable MP retention even under pore-scale flow. This is consistent with nano- and microscale measurements using CP-AFM and Raman imaging, where eco-corona bridging forces far exceed environmental separational forces. Moreover, pore-scale flow simulations show that separational forces under our experimental conditions are insufficient to disrupt this retention (Figure S7). Relative particle velocity within pores further modulates interaction strength. Once MPs enter the pore space, they stay there until local separational forces exceed the eco-corona bridging forces. This means depositional zones store MPs during base flow but can release them during floods or dredging. Mechanically trapped MPs, governed by pore geometry and particle size, remain largely immobilized within the sediment matrix. Over time, surface weathering, abrasion and eventually fragmentation can create smaller, surface-altered MPs. These smaller MP particles are likely even more susceptible to eco-corona formation, due to their higher surface-to-volume ration. Consequently, this reinforces aggregation and promotes long-term retention. This mechanism explains how originally buoyant or even low-density MPs can overcome buoyancy constraints and accumulate in sediments. With sediments acting as a persistent environmental sink for MP. The long-term fate of MPs in aquatic systems is shaped by sediment-mediated transport and downstream deposition.

## Detailed understanding of eco-corona mediated attraction

Up to this point, we showed and tested eco-corona bridging forces across length scales inducing heteroaggregation and subsequent sedimentation. In the environment, the ionic strength varies significantly from freshwater to saltwater systems and sediment particles are also highly likely to be covered by an eco-corona. Therefore, we investigated these two major parameters influencing the heteroaggregation by CP-AFM: eco-corona coverage and ionic strength. With this we identify the parameter space in which the bridging mechanism occurs.

### Eco-corona coverage influences bridging forces between MP and sediment particles

With increasing eco-corona coverage in the interaction area of MPs and the colloidal model sediment ($SiO_2$) probe the eco-corona bridging force decreases. The bridging force decreased stepwise from each measured particle to the next particle (Figure 3A and B, Figure S12). When the sediment colloidal probe approaches and contacts an eco-corona covered MP particle, macromolecules of the eco-corona are transferred during retraction. This increases eco-corona coverage and reduces the number of free adsorption sites as the number of measured particles (n) increases. The eco-corona bridging forces during approach and retraction are subsequently decreasing. Figure 3D summarises this mechanism. The observed decrease in bridging force following macromolecule transfer, agrees with findings of

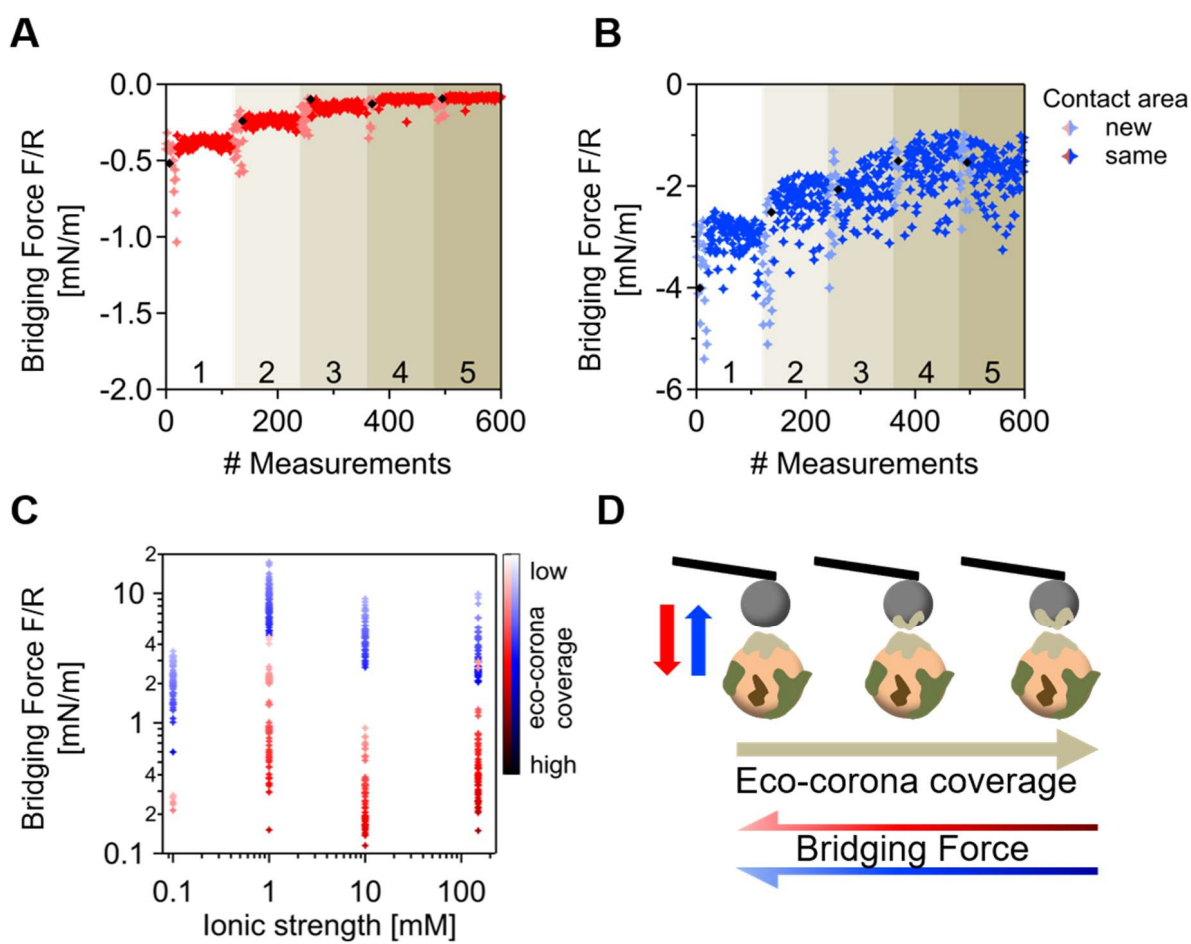

*Figure 3: Influence of eco-corona coverage (A, B) and combined effects of coverage and ionic strength (C) on eco-corona bridging forces between $SiO_2$ and PS particle. Bridging forces on approach (red) and retraction (blue). (A) Bridging forces are dependent on the eco-corona coverage in the contact area. With new particle areas measured the eco-corona coverage on the $SiO_2$ probe increases. 20 different contact areas (bright colour) were measured on 5 different PS particles (beige gradient). The black marker highlights the contact area within the force maps (Figure S12D) where 100 consecutive measurements in the same contact area (dark colour) were performed. PS incubated for 2 weeks in freshwater and measured in 150 mM KCl. Bridging forces are highest for lowest eco-corona coverage. Bridging forces decrease with increasing eco-corona coverage up to a point where no bridging forces are detected. (C) Bridging forces on approach (red) and retraction (blue) at different ionic strengths. Shade of the colours indicates the amount of eco-corona coverage. From bright, i.e. low coverage to dark, i.e. high coverage. Bridging forces measured at different ionic strength and surface coverage all exceed the simulated separational force of $4 \cdot 10^{-4}$ mN/m. (D) Sketch of the bridging force in dependence of eco-corona particle coverage.*

Klein et al. who thoroughly investigated polymer bridging interactions under various conditions.[70-72] The repulsive interaction range is only dependent on the chain length of the bridging macromolecules and not on the number of free adsorption sites. For the repulsive interaction range on retraction $x_1$ and approach $x_2$ < 50 nm bridging is always present (Figure S13). For $x_1$, $x_2$ > 50 nm bridging was never observed. Next to the bridging force's magnitude, these two parameters can act as proxies for the degree of eco-corona coverage.

Sun et al. found that the amount of adsorbed humic acid strongly influences the aggregation rate of PS nanoparticles.[73] The aggregation rate decreased with higher concentrations of humic acid due to steric stabilisation of the particles. The steric stabilisation is the result of decreasing numbers of free adsorption sites at the particles surfaces, reducing bridging forces as described above. This shows that our proposed mechanism likely applies to nanoplastics as well. In addition, our results suggest quasi time-dependent changes in particle properties, since eco-corona formation is an equilibrium process. Longer exposure time of MP to NOM and higher NOM concentrations are likely to increase eco-corona coverage until equilibrium is achieved. Thus, reducing bridging and heteroaggregation.

The interaction behaviour of eco-corona or biofilm covered MP is dominated by the properties and the adsorption density of the macromolecules in the interaction area. Importantly, bridging interaction occurs when at least one surface has low macromolecule coverage, while purely repulsive interaction occurs between two highly covered surfaces as we have shown in previous work.[28] The coverage of MP is thereby dependent on the NOM type, concentration and exposure time. In addition, partial coverage may be the result of particle abrasion, eco-corona/biofilm abrasion or MP particle disintegration.[23,24,74-76]

### Ionic strength influences eco-corona bridging between MP and sediment particles

In addition to the eco-corona coverage, ionic strength also effects the eco-corona bridging force between MP and sediment particles, influencing aggregation behaviour and sediment retention. In natural waters, MP particles are exposed to a wide range of ionic strengths and ion composition, which can alter their interactions.

At 0.1 mM KCl, representing model rainwater, both bridging forces are smallest during approach and retraction (Figure 3C). Electrostatic repulsion dominates under these conditions, suppressing eco-corona bridging. At 1 mM KCl, representing model freshwater, bridging forces reach their maximum. At ionic strengths, comparable to model brackish water (10 mM KCl) and model saltwater (150 mM KCl) bridging forces were similar, but both were lower than those at 1 mM. This pattern suggests the ionic strength has two opposing effects leading to a peak in bridging forces at intermediate concentrations (1 mM KCl).

The ionic strength affects the bridging forces by influencing (1) electrostatic interactions, (2) hydrogen bonding, and (3) conformational changes of the macromolecular chains (Figure S1 and Figure S13). At low ionic strength (0.1 mM, model rainwater), the Debye length exceeds the typical interchain distance of adsorbed polyelectrolytes (4 -11 nm),[77,78] preventing charge screening. This causes macromolecular chains to extend due to intra- and interchain repulsion, which limits their ability to rotate and bind to free sites. In addition, for typical environmental pH-values (pH > 3) the surface of the sediment particle is negatively charged.[79] Therefore, adding an electrostatic repulsive force, since we have shown in previous work that the zeta-potential of eco-corona covered MP is negative as well. At 1 mM KCl electrostatic repulsive forces are screened to length scales of interchain distances. Thereby, drastically reducing the repulsive interaction range on approach (Figure S13A). The maximum bridging forces at 1 mM KCl (freshwater) occur due to an interplay of reduced repulsive forces and increasing attraction. Attractive forces are possibly present from eco-corona functional groups that are positively charged, e.g. amines, or able to form hydrogen bonds, e.g. hydroxyls, carboxyls, to attach to the sediment surface.[80] Local positive charges within a NOM macromolecule chain could cause electrostatic attraction to negatively charged sediment, although the overall charge of the eco-corona is negative. Besides electrostatic attraction, hydroxyls and carboxyls could form hydrogen bonds to the surface of the sediment, promoting bridging. At 10 and 150 mM KCl, the increased ionic strength is screening any electrostatic interaction to a minimum (Figure S13). In addition, hydrogen bonds are also weakened,[81] decreasing bridging forces (Figure 3C). At this level of ionic strength, the eco-corona coverage becomes a stronger determinant for bridging than ionic strength.

Multivalent cations can also enhance eco-corona bridging between negatively charged surfaces by reducing or inverting surface charge, as supported by aggregation experiments and molecular dynamics simulations.[43,73,82] Zeng et al. showed that increasing calcium cation concentrations promote nanoplastic-goethite heteroaggregation in the presence of humic acid.[83] Furthermore, Hakim and Kobayashi found that polyethylene particles, in the presence of natural organic matter, formed stronger aggregates in dispersions with calcium cations (0.6–3.8 mN/m) than with potassium cations (0.4–1.4 mN/m).[36] Although our measured bridging forces (1–17 mN/m) are within a comparable range, the higher values may only reflect differences in eco-corona composition, or measurement methods. Nevertheless, the strong bridging forces allow aggregates to be stable in the environment from low to high salt concentrations, especially in the presence of multivalent cations. Even fast flowing currents cannot disintegrate these aggregates as our simulations demonstrated (4 · $10^{-4}$ mN/m, Figure S2).

## Conclusion

Eco-corona covered MP can show attractive bridging forces with sediment particles under specific circumstances, in contrast to plain MP or in the absence of sediment particles. In these interactions, the eco-corona largely dominates the interaction, while the surface properties of the originally plain MP have a minor influence. We experimentally identified an eco-corona bridging mediated aggregation mechanism between MP and sediment particles on different length scales (nano, micro and macro). This gives a mechanistic explanation for MP–sediment heteroaggregation. The degree of eco-corona coverage in the contact area of the interacting particles mainly influences the strength of the bridging force. The mechanism requires one surface to be only partially covered with eco-corona. If both surfaces are fully covered, the interaction becomes repulsive as we have shown in previous work. Partial eco-corona coverage may result from abrasion or particle disintegration. This heteroaggregation mechanism appears relevant in all aqueous environments, except low ionic strength waters such as rainwater (≤ 0.1 mM). In our model rainwater, the eco-corona mediated interaction is repulsive. In aquatic environments with ionic strengths of at least 1 mM, i.e. freshwater, brackish water or saltwater (1, 10, 150 mM) the interaction becomes attractive, making eco-corona bridging mediated heteroaggregation the dominant interaction mechanism for MP. Our simulations demonstrated that MP heteroaggregates form and remain stable even under environmentally high separational force conditions. Except in very fast flows, this mechanism is expected to promote MP-sediment heteroaggregation and eventually sedimentation through increased overall aggregate density. This mechanism is valid across sediments of different compositions and MPs of diverse polymer types when exposed to natural organic matter (NOM) and microorganisms, which we have demonstrated on biofilm covered, low-density MP interacting with bentonite clay. Following this mechanism, sedimentation occurs in the water column when suspended sediment concentrations are sufficient for MPs to cross the buoyancy barrier. Once sedimented, MPs infiltrate sediment beds, where the presence of an eco-corona enhances their retention. Consequently, calm riverbeds and lake floors can serve as major environmental sinks for MPs. Effective risk assessment must therefore account for MP transport and accumulation dynamics. In fast-flowing waters, dispersion increases exposure risk to organisms in the water column. But in stagnant zones, sedimentation promotes long-term contamination and exposure to organisms in sediments. Recognising these contrasting dynamics is essential for targeted monitoring and risk evaluation.

To conclude, the eco-corona- and biofilm-mediated bridging mechanism induces MP-sediment heteroaggregation which leads to an increase in overall density. Subsequently, even low-density MP crosses the buoyancy barrier and sediments. Eventually, MP accumulation zones will be found at and within sediments of calm parts of water bodies and sediment of flood plains as they are retained here and won't get easily resuspended by water.

# Material and Methods

## Microplastic particle incubation

PS particles with a diameter of 3 µm (Prod. Nr. 01-00-303, Micromod Partikeltechnologie GmbH, Rostock, Germany), PS polymer pellets (PS 158N, INEOS Styrolution Group, Frankfurt am Main, Germany) and Rhodamine-B (RhB) labelled PS pellets (950904, Magic Pyramid, Frechen, Germany) were purchased. PS pellets were cryo-milled with an ultra-centrifugal mill (ZM-200, Retsch GmbH, Haan, Germany) with a 12 teeth rotor and a 200 µm distance sieve. Particle fractions $d_{90}<20$ µm and $d_{90}=20-75$ µm were separated using an Alpine air jet sieve (E200 LS, Hosokawa Alpine AG, Augsburg, Germany).

PS particles were incubated using either basin freshwater, for eco-corona development, or ultrapure water, as control, following the procedure of Ramsperger et al.[31] In short, 1) 100 µL of the 3 µm PS MP particle suspension, used for nanoscale experiments, was given into 900 µL of corresponding medium. 2) 200 mg of $d_{90}<20$-75 µm PS MP particles, used for microscale experiments, were incubated with 20 mL of the corresponding medium. 3) 100 mg of $d_{90}<20$ µm RhB-PS (RhB-PS, cryo-milled, <20 µm, FW44, Z01, SFB 135, more information in SI), used for macroscale experiments, were incubated in 20 mL of the corresponding media. To promote surface interactions and prevent sedimentation, all vials were placed on a sample roller. The basin freshwater was replaced three times per week to ensure a continuous supply of natural organic matter (NOM) and active microbial communities. For medium exchange, suspensions were centrifuged for 20 min at 2000 g, and 900 µL or 18 mL of the supernatant (depending on the MP particles size) was carefully replaced with freshwater or ultrapure water. The MP particles were incubated for 2 weeks and thereafter stored at 8 °C until further analysis in order to minimize microbial growth. The basin freshwater was analysed (details are provided in the SI), with the following parameters: total organic carbon (TOC) = 8.8 mg/L, pH = 8.45, and cationic contribution to ionic strength ≈ 5.3 mM, determined via inductively coupled plasma optical emission spectroscopy (ICP-OES).

MP particles used in macroscale aggregation and sedimentation experiments, were produced from low-density polyethylene (LDPE, MKCP2615 / Sigma-Aldrich, USA). LDPE particles were produced by cryogenic grinding using the device Pulverisette 0 (Fritsch, Idar-Oberstein, Germany) and liquid nitrogen. To obtain the targeted size range of 63 – 200 µm the MP particles were sieved with the vibratory sieve shaker (AS 200, Retsch, Germany). For biofilm formation, river water was sampled at the Elbe River (GPS coordinates: 51.052083, 13.812333 decimal degree [N, E]) and filtered with a paper filter (10 – 15 µm, VEB Freiberger Zellstoff- und Papierfabrik, Germany). The water was enriched with 1.0 g/L glucose and eight drops of fertiliser (indoor palm tree fertiliser, Gebr. Mayer Produktios- und Vertriebsgesellschaft mbH, Germany). Subsequently, 1 g of LDPE particles was added to 500 mL river water in sterile 1 L Erlenmeyer flasks. Subsequently, flasks were covered with aluminium foil and incubated at 30 °C and 70 rpm for two weeks in an incubator shaker (Innova S44i, Eppendorf, Germany). During incubation a day-night-cycle of 12 h was applied. After the incubation, particles were separated from the incubation media using a 63 µm mesh (Fritsch, Germany) and washed with ethanol on a paper filter (10 – 15 µm, VEB Freiberger Zellstoff- und Papierfabrik, Germany). The washed particles were stored in ethanol for one week to sterilise the samples, filtered and air-dried before further use, while preserving the biofilm (Figure S6).

## Colloidal Probe-Atomic Force Microscopy (CP-AFM)

Colloidal Probe direct force measurements (Figure 1A) were conducted using an MFP-3D Bio AFM (Asylum Research Inc., Santa Barbara, USA). Tipless cantilevers (CSC37B and CSC38A, Mikromasch) were calibrated using the thermal noise method.[84,85] The spring constants of CSC37B were 0.2 – 0.3 N/m and 0.02 – 0.03 N/m for CSC38A.

The preparation of the colloidal probes and measurement procedure followed the description of Witzmann et al.[28] The cantilevers were rinsed with ultrapure water, ethanol, ultrapure water and iso-propanol to remove any debris. Meanwhile, silica ($SiO_2$) sediment colloidal particles (nominal diameter of 1.76 µm, microParticles GmbH, Berlin, Germany) were washed in ethanol and glued to the cantilever. For this 2-component epoxy glue (UHU Plus Endfest, UHU GmbH & Co. KG, Bühl/Baden, Germany) was applied to the cantilever with a micro capillary. The glue was cured between 60 – 80 °C for at least 5 h. The substrates were modified with polydimethylsiloxane (PDMS) to immobilise MP particles.[86-88] Four different potassium chloride (KCl) concentrations, 0.1 mM, 1 mM, 10 mM and 150 mM were used to mimic rainwater, freshwater, brackish water and saltwater in our experiments.

### Data evaluation of CP-AFM experiments

AFM raw data was converted to force separation curves via an updated version of the IGOR Pro procedure of M. Seuss.[89] This allowed batch conversion of the raw data into force separation curves. After obtaining force separation curves for the approach and retraction they were normalised by applying the Derjaguin approximation (eq. $\frac{F}{R_{eff}} = 2\pi W$,

(4)).[90] This allows comparing interaction forces of surfaces with different geometries.

$$\frac{F}{R_{eff}} = 2\pi W, \quad (4)$$

with $W$ being the interaction energy and where the effective radius $R_{eff}$ is given by:

$$R_{eff} = \frac{R_1 R_2}{R_1 + R_2}. \quad (5)$$

$R_1$ and $R_2$ being the radii of the two interacting particles, e.g., model sediment and eco-corona covered MP particle. The following data-analysis steps were performed on the obtained force/effective radius separation curves in Figure S11. (1) Running a median-smoothing within 7 data points, (2) determining the force minimum, i.e. attractive force, (3) finding zero crossings of force $F = 0$ by the change of the sign starting at the force minimum and from there going left and right. We excluded data points from the interaction range graphs where attractive (negative) forces never exceeded the noise level and as such the procedure failed to determine a correct value for the minimum force or zero crossing.

### Numerical simulation of environmental separation force

Numerical simulations were performed using the Lattice-Boltzmann method. Two fixed spheres with diameters 3 µm (MP) and 1.8 µm (sediment) were placed into a cubic computational box with an edge length of 9 µm (Figure 1B). The distance between the surfaces of these spheres was 0.8 µm. A homogeneous shear flow was generated by imposing a constant velocity at the top and bottom walls, which exerted a separating force on the spheres. To achieve faster convergence, we run simulations at higher shear rates. The resulting force was linearly scaled down to realistic environmental shear rates, which is allowed due to the low Reynolds number (Stokes regime). To estimate a realistic shear rate $\dot{\gamma}$, we started with the turbulent kinetic energy $E$ which is in the range of $10^{-5}$ W/kg[91] in typical rivers.

$$\dot{\gamma} = \sqrt{E\frac{\rho}{\eta}} \quad (6)$$

with $\rho$ and $\eta$ denoting the density and viscosity of water at ambient conditions, respectively. This leads to a shear rate of $\dot{\gamma}$ = 3.1 1/s.

To represent a porous environment n silica spheres (n = 0, 1, 2, 3, 4, and infinite) with a diameter of 3 µm were arranged such that the surface-to-surface distance between the microplastic (1.8 µm) and silica spheres was 0.1 µm. A uniform flow field was generated by imposing a constant velocity (u = 1 m/s) at the inlet and outlet boundaries, generating a separation force on the microplastic sphere. The remaining boundaries of the domain were treated as periodic. The resulting separation force scaled linearly with the imposed flow velocity, consistent with the low Reynolds number regime (Stokes flow). For consistent comparison across cases, the ratio of separation force to effective radius was evaluated.

### Aggregation experiments of PS and model sediment particles

Heteroaggregation experiments have been performed with plain and eco-corona covered PS particles (20 - 75 µm). The model sediment stock dispersion (1.67 · $10^{23}$ particles/mL, diameter 2.83 µm, microParticles GmbH, Berlin, Germany) was 10-fold diluted in 1 mM KCl. 400 µL of this dilution were added to a 1.5 mL glass vial. 200 µL of the eco-corona covered PS suspension was infused into the model sediment dispersion (Figure 1C). Then, the vial was gently rolled across the lab bench at an estimated relative velocity of 1 mm/s. The PS particles were allowed to settle for a minute and 400 µL of the supernatant was replaced with 1 mM KCl solution. The rolling, sedimentation and dispersion replacement was repeated. For the following two repetitions 800 µL KCl were added and replaced. Resulting in a 225-fold dilution of the unbound model sediment particles. During this procedure, model sediment and PS particles aggregates formed. These aggregates were subsequently analysed by optical microscopy and µRaman spectroscopy (Figure 1C).

### Raman imaging of MP-model sediment aggregates

To differentiate materials within the heteroaggregates, µRaman spectroscopy was performed using a confocal µRaman setup (WITec alpha 300 RA+ Raman imaging system), equipped with a UHTS 300 spectrometer and a back-illuminated Andor Newton 970 EMCCD camera, controlled by the WITec Suite SIX 6.1 software. Excitation was provided by a 532 nm laser at a power of 20 mW for all measurements. A drop of the aggregate dispersion ($V \approx 0.2$ mL) was placed on a circular coverslip mounted in an in-house-made aluminium support with a circular opening of $d = 1$ cm, providing enough space to access the droplet with a water immersion objective (Zeiss W N-Achroplan 63x / *NA* = 0.9). This way, the drop was entrapped reducing undesired particle motion during measurement due to evaporation. For Raman imaging a lateral step size of 0.5 µm/pixel and an integration time of 0.5 s/pixel were used. The spatial component distributions were extracted from the imaging data using the True Component Analysis tool embedded in the WITec Project SIX+ software. Raman spectra at specific locations within the aggregates, as well as spectra of neat components (dry PS particles and model sediment particles suspended in water, Figure S4), were acquired with an integration time of 0.5 s and 50 accumulations. All spectra were processed with cosmic ray removal and baseline correction routines.

### Aggregation-Sedimentation Experiments with LDPE MP Particles

The aggregation-sedimentation experiments (Figure 1C) were carried out with a flocculation tester (WiseStir JT-M6C, witeg Labortechnik GmbH, Germany) in 800 mL glass beakers. For all experiments, Elbe River water was used after filtration with a paper filter (10 – 15 µm, VEB Freiberger Zellstoff- und Papierfabrik, Germany). The settling ratio of pristine and biofilm covered LDPE was determined in the presence and absence of bentonite according to the parameters in Table 1.

*Table 1: Parameters of sedimentation experiments to investigate the influence of biofouling on the settling ratio.*

| Name | $c_{MP}$ in g/L | $c_{Ben}$ in g/L | Biofilm |
|---|---|---|---|
| LDPE + $H_2O$ | 0.125 | 0 | No |
| LDPE + bentonite | 0.125 | 0.625 | No |
| LDPE + biofilm | 0.125 | 0 | Yes |
| LDPE + bentonite + biofilm | 0.125 | 0.625 | Yes |

In the first step, bentonite (70 % montmorillonite, Roth GmbH, Germany) $d_{90}$ = 35.95 ± 0.68 µm (Table S5) was added to the water and mixed until fully suspended. Afterwards, the stirring speed was set to 200 rpm, and LDPE particles were added. Each experiment was run for 120 min. After stirring, the agitators were removed and attached particles were carefully rinsed back into the beaker using deionised water. Then, the samples were covered with aluminium foil and left overnight for the LDPE and bentonite particles to settle. The floating LDPE particles were removed and collected by skimming and overflowing of the beaker with 400 mL water. The collected particle suspensions were filtered with a 10 – 15 µm paper filter. Lastly, the mass of the floating LDPE particles was determined after drying at 60 °C and used to calculate the settling ratio:

$$settling\ ratio = \frac{m_{LDPE,0} - m_{LDPE,floating}}{m_{LDPE,0}} \cdot 100\%, \qquad (7)$$

where $m_{LDPE,0}$ is the initial mass of the added LDPE particles, $m_{LDPE,floating}$ is the remaining mass of floating LDPE after the experiment.

## Comparative Column Transport Experiments of Plain and Eco-corona covered MP

Plain and eco-corona covered RhB-PS were prepared as 200 mg/L suspensions in 1 mM KCl. The 1 mM KCl was chosen to maximise eco-corona bridging forces, as supported by the findings presented in the section on the Ionic strength influences eco-corona bridging between MP and sedimentIonic strength influences eco-corona bridging between MP and sediment . There are two types of MP loading suspensions for this column transport experiment: A. Eco-corona–covered RhB-labelled PS (Eco-RhB-PS); B. plain RhB-labelled PS (Plain-RhB-PS). The matrix of this column experiments used is quartz sand (FH31, Quarzwerke GmbH), that had been heated at 350 °C for 8 hours to remove any potential organic matter. The quartz sand had a particle density of 2.65 g/cm³, a bulk density of 1.7 g/cm³, and a porosity of 45% and the particle size distribution consisted of 23.3% coarse sand, 57.0% medium sand, and 19.7% fine sand (more details are provided in Figure S13). For each MP loads, three quartz sand filled columns were used in parallel to obtain drainage data suitable for statistical analysis. Therefore, 6 columns (inner diameter: 10.3 mm; inner height: 74.3 mm; volume: 6.19 cm³, workshop of the University of Bayreuth) were packed with the pre-heated quartz sand. Ceramic filters (pore size: 200 µm, workshop of the University of Bayreuth) were fitted at both ends of each column to prevent sand from entering the pump tubing and blocking the flow. The columns were connected to calibrate peristaltic pumps (12-channel Masterflex L/S Pump System, Cole-Parmer, USA) using tubing with an inner diameter of 0.18 mm to ensure a constant flow rate.

In this column transport experiments, we selected pumping rates of 0.08 mL/min and 0.25 mL/min to represent typical environmental hydraulic conditions. The average pore flow velocity

$$u_{avg} = \frac{Q}{A\epsilon} \qquad (8)$$

inside the column is determined from the pumping rate $Q$ divided by the column cross-sectional area

$$A = \pi \frac{D^2}{4}. \qquad (9)$$

Given the column's porosity $\epsilon = 0.45$ the average pore velocity for 0.08 mL/min is $u_{0.08}$ = 3.56 · 10⁻⁵ m/s and $u_{0.25}$ = 1.11 · 10⁻⁴ m/s for pumping rates of 0.25 mL/min, respectively. The Reynolds number is used to assess the flow regime, calculated as

$$Re = \frac{\rho u_{avg} d_{50}}{\mu} \qquad (10)$$

where $\rho = 998\ kg/m^3$ is the density of water at 20°C, µ = 1.003 · 10⁻³ Pa·s is the 1mM KCl dynamic viscosity, and $d_{50}$ = 390.8 µm is the median grain diameter of the quartz sand. The resulting Reynolds numbers are: for pumping rates of 0.08 mL/min, $Re_{0.08} \approx 0.014$, and for pumping rates of 0.25 mL/min, $Re_{0.25} \approx 0.043$. These values indicate laminar flow, where microplastic particle transport is dominated by advection and diffusion, with negligible inertial effects.

To avoid potential interference from dispersants, the experiment followed the dispersant-free column design displayed in Figure S15. This setup used MP loading suspensions in large-base petri dishes, keeping the liquid height as low as possible without changing the total volume and applying continuous stirring to minimise the effect of the plain MPs' inherent buoyancy. In addition, each column was paired with a blank reference channel (BL), with both inlet tubes fixed together to ensure sampling from the same liquid position. This allowed precise monitoring of the MPs' amount introduced into each column. Drainages were collected in glass sample bottles to eliminate potential MP contamination during subsequent quantification.

There were four sequential flow phases in these experiments: (1) In the Saturation Phase (S), all quartz columns were saturated with 1 mM KCl at 0.08 mL/min for 80 minutes (6.40 mL per column, 1.91 pore volumes) to remove air pockets and establish stable saturation, with drainages collected as background reference samples. (2) In the MP Loading Phase (L), each MP suspension was introduced at the same flow rate (0.08 mL/min) for 80 minutes, producing 6.40 mL of drainage per column (1.91 pore volumes). The same applies to each parallel paired blank reference channel (BL). (3) In the Flushing Phase (F), the columns were flushed with 1 mM KCl at 0.08 mL/min for 160 minutes (12.8 mL, 3.8 pore volumes) to assess the detachment of loosely bound particles under low to moderate hydraulic forces. (4) In the Super-Flushing Phase (SF), the same background solution was applied at 0.25 mL/min for 32 minutes (8.0 mL, 2.4 pore volumes) to simulate higher shear stress (moderate-high flow conditions) and remove particles not retained by size exclusion.

Subsequently, RhB-PS particles in drainage were quantified using a fluorescence microscope with hemocytometers. This method was validated for reliability and reproducibility and calibrated with a linear curve ($R^2 = 0.99$). Three fluorescence filters were used: DTAF (492/516 nm), Cy3 (554/568 nm), and DAPI (350/465 nm). Drainages from the saturation phase (W) were first examined for background fluorescence from quartz sand, which was negligible under the Cy3 filter. Each drainage was mixed with 0.25% (v/v) Tween 20, shaken overnight to ensure uniform dispersion, and 10 µL loaded into a hemocytometer (0.1 mm depth, 0.0025 mm² grid, Improved Neubauer, Hecht Glaswarenfabrik GmbH & Co KG), avoiding air bubbles or overflow. Samples were then examined under a fluorescence microscope (Axiostar Plus, Zeiss) with a Plan-NEOFLUAR 10x/0.30 objective and Cy3 filter set (554/568 nm), illuminated by a constant HXP 120 light source (LEJ, Leistungselektronik JENA GmbH). Particles were counted in the four corner grids and the central grid, including only intact, clearly distinguishable fluorescent particles. Aggregates were analysed using 3D topographic imaging to determine the number of RhB-PS particles, and the standard two-side inclusion rule was applied for particles touching grid borders. Each sample was analysed in duplicate, and counts were repeated if replicate measurements differed by more than ±5%.

Each retention was calculated as the percentage of MP particles retained in the transport column relative to the total number introduced during the MP loading phase. Mathematically, retention at each stage was defined as:

$$\text{Retention (\%)} = \left(1 - \frac{N_{out}}{N_{in}}\right) \cdot 100 \tag{11}$$

where $N_{in}$ is the number of RhB-PS particles introduced into the column during the MP loading phase (L), and $N_{out}$ is the number of particles detected in the drainage of each phase (L, F, SF).

Subsequently, we performed statistical analysis on the obtained data using DABEST (Data Analysis with Bootstrap-coupled ESTimation)[2]. This method focuses on estimation rather than relying solely on hypothesis testing and provides a more informative perspective than traditional significance tests such as p-values. By using bootstrap resampling, we quantify the magnitude and precision of differences between groups, making the results easier to interpret in a practical context.

In this experiment, each MP load had three parallel columns, and each drainage was analysed in duplicate to ensure reproducibility, so the experiment data is independent and identically distributed, which meets the requirements of DABEST. We applied 5000 bootstrap resamples to compare microplastic retention between eco-corona-covered RhB-labelled PS (Eco-RhB-PS) and plain RhB-labelled PS (Plain-RhB-PS). This allowed to obtain robust estimates and reliable confidence intervals without relying on strict parametric assumptions. The resulting Gardner–Altman plots and unpaired mean differences with 95% bias-corrected and accelerated confidence intervals are shown in Figure S8A and B and Figure S9.

# Author Information

**Corresponding Authors:**

Andreas Fery – Leibniz Institute of Polymer Research Dresden e.V., Institute of Physical Chemistry and Polymer Physics, 01069 Dresden, Germany; Physical Chemistry of Polymeric Materials, Technische Universität Dresden, 01069 Dresden, Germany; orcid.org/0000-0001-6692-3762; Email: fery@ipfdd.de

Günter K. Auernhammer – Leibniz Institute of Polymer Research Dresden e.V., Institute of Physical Chemistry and Polymer Physics, 01069 Dresden, Germany; orcid.org/ 0000-0003-1515-0143; Email: auernhammer@ipfdd.de

**Acknowledgments**

The authors acknowledge funding from the Deutsche Forschungsgemeinschaft (DFG; German Research Foundation) project number 391977956–SFB 1357 Microplastics and 422852551 within the priority program 2171. We acknowledge Johanna Schmidtmann for DOC/TOC-analysis and Christine Steinbach for the cation analysis with ICP and anion analysis with UV/VIS. We thank the keylab "Synthesis and Molecular Characterization" of the Bavarian Polymer Institute (BPI) at the University of Bayreuth for support. ChatGPT was used in parts to improve wording and readability.